\documentclass{PoS}

\title{Combination and QCD Analysis of the HERA Inclusive Cross Sections}

\ShortTitle{QCD Analysis of the HERA Inclusive Cross Sections}

\author{\speaker{Voica Radescu}\thanks{On behalf of the H1 and ZEUS collaborations}\\
Physikalisches Institut, University of Heidelberg\\
        E-mail: \email{voica@mail.desy.de}}

\abstract{
A QCD fit analysis to the combined HERA-I inclusive deep inelastic cross sections measured by the H1 and ZEUS collaborations for $e^\pm p$  scattering resulting into a competitive NLO PDF set, HERAPDF1.0 is presented. 
HERAPDF at NNLO fits are presented as well, resulting, however, in a worse description of the combined HERA data.
In addition, a preliminary  analysis including the HERA II measurements of lower proton-beam energies
 is performed. The effect of including the new data on the determination of HERA parton
distribution functions is analysed, using fits similar to those performed for HERAPDF1.0. Some tension of the QCD 
fit with respect to the data is identified in the kinematic region of low $Q^2$ and low x. Finally, the QCD fit analysis of the combined HERA-I inclusive deep inelastic cross sections has been extended to include 
combined HERA II measurements at high $Q^2$. The effect of including these data on the determination of parton 
distribution functions is analysed, resulting into HERAPDF1.5. The precision of the new 
PDFs at high-x is considerably improved, particularly in the valence sector.}

\FullConference{35th International Conference of High Energy Physics - ICHEP2010,\\
		July 22-28, 2010\\
		Paris France}

\begin{document}

%%%%%%%%%%%%%%%%%%%%%%%%%%%%%%%%%%%%%%%%%%%%%%%%%%%%%%%
%%%%%%%%%%%%%%%%%%%%%%%%%%%%%%%%%%%%%%%%%%%%%%%%%%%%%%%
%%%%%%%%%%%%%%%%%%%%%%%%%%%%%%%%%%%%%%%%%%%%%%%%%%%%%%%

\section{Combined H1 and ZEUS Cross Section Measurements }\label{sec:data}

\vspace*{-0.3cm}
Deep inelastic scattering (DIS) at HERA has been crucial to the exploration
of proton structure and quark-gluon interaction dynamics.  
The combination of the H1 and ZEUS data provides the most accurate measurements of DIS inclusive double differential cross-sections of neutral (NC) and charged current (CC)  $e^\pm p$ scattering over an extended  kinematic range of  $0.045 < Q^2 < 30\,000$~GeV$^2$ and $6\times10^{-5}<x<0.65$. Therefore, these
accurate measurements can be used as a sole input to the QCD analysis to determine the proton structure - parton distribution functions
(PDFs) as described in the following sections, which can be used then for precise predictions for the LHC processes.

The combination of data uses the $\chi^2$ minimisation method and it takes into account the correlated systematic uncertainties
for the H1 and ZEUS cross-section measurements \cite{h1pdf}. 
This combination procedure is applied for the following scenarios:
to the published inclusive deep inelastic cross sections measured by the H1 and ZEUS 
collaborations in CC and NC unpolarised $ep$ scattering at HERA during the period 1994-2000, termed HERA I;
%which used into HERA QCD anayielded the HERAPDF1.0  \cite{herapdf};
to the preliminary inclusive deep inelastic cross sections measured by the H1 and ZEUS 
collaborations in NC unpolarised $ep$ scattering at HERA during its last months of operation in 2007 with reduced
proton beam of  $E_p=460 GeV$ and $E_p=575 GeV$ which provides additional PDF constraint in the low $Q^2$, low $x$ region; and to the preliminary  inclusive deep inelastic NC and CC cross sections at high $Q^2$ measured by the H1 and ZEUS 
collaborations at HERA during the whole period which considerably improve PDF uncertainties at high-$x$
as described in these proceedings. 
%%%%%%%%%%%%%%%%%%%%%%%%%%%%%%%%%%%%%%%%%%%%%%%%%%%%%%%
%%%%%%%%%%%%%%%%%%%%%%%%%%%%%%%%%%%%%%%%%%%%%%%%%%%%%%%
%%%%%%%%%%%%%%%%%%%%%%%%%%%%%%%%%%%%%%%%%%%%%%%%%%%%%%%

\section{QCD Analysis settings}\label{sec:analysis}

\vspace*{-0.3cm}
The above combined data is used as a sole input into a QCD fit analysis to extract the proton's PDFs.
% therefore these proceedings present three types of 
%fit results: finalised results based on HERA I data (HERAPDF1.0), preliminary results reflecting the effects of including the 
%low energy proton data runs in addition to HERA I data, and preliminary fit results based on combined HERA I and II data (HERAPDF1.5).
The HERA data have a minimum invariant mass of the hadronic system, $W$, of $15$ GeV and
a maximum $x$ of $0.65$, such that they are in a kinematic region where there is no sensitivity to
target mass and large-$x$ higher-twist contributions. 
In addition, to restrain to the region where perturbative QCD is valid, only data above $Q^2_{min}=3.5$~GeV$^2$ is used in the central fit.
%, with its variation being studied.

The fit procedure consists first in parametrising PDFs at a starting scale  $Q^2_0=1.9~ \rm GeV^2$,
chosen to be below the charm mass threshold.
 The parametrised PDFs are the valence distributions
 $xu_v$ and  $xd_v$,  the gluon distribution $xg$, and the $u$-type and $d$-type 
$x\bar{U}$, $x\bar{D}$, where $x\bar{U} = x\bar{u}$, 
$x\bar{D} = x\bar{d} +x\bar{s}$. 
The following standard functional form is used to parametrise them
\begin{equation}
 xf(x) = A x^{B} (1-x)^{C} (1 + D x + E x^2),
\label{eqn:pdf}
\end{equation}
where the normalisation parameters, $A_{uv}, A_{dv}, A_g$,  are constrained by  
the QCD sum-rules. 
The $B$ parameters  $B_{\bar{U}}$ and $B_{\bar{D}}$ are set equal,
 $B_{\bar{U}}=B_{\bar{D}}$, such that 
there is a single $B$ parameter for the sea distributions. 
The strange quark distribution 
is already present at the starting scale and 
it is  assumed here that 
$x\bar{s}= f_s  x\bar{D}$ at $Q^2_0$. 
The  strange fraction is chosen to be $f_s=0.31$ which is
consistent with determinations 
of this fraction using neutrino induced di-muon production. 
In addition, to ensure that $x\bar{u} \to x\bar{d}$ 
as $x \to 0$,  
$A_{\bar{U}}=A_{\bar{D}} (1-f_s)$.
The $D$ and $E$ are introduced one by one until further improvement in $\chi^2$ is found.
The best fit  results in a total of 10 free parameters.

The PDFs are then evolved using DGLAP evolution equations \cite{qcdnum}.  at NLO  and NNLO in the $\overline{MS}$ scheme with the
renormalisation and factorisation scales set to $Q^2$.
% and the world average value for the strong coupling 
%$\alpha_s(M_Z) =  0.1176$~\cite{pdg}. 
%In these proceedings, also fits at NNLO are presented for different values of strong running coupling,  $\alpha_s(M_Z) =  0.1176$ and $\alpha_s(M_Z) =  0.1145$.
The QCD predictions for the structure functions 
are obtained by convoluting 
the PDFs with the calculable coefficient functions taking into account mass effect for the heavy quarks based on the general mass variable flavour scheme \cite{tr}.
%
%In this analysis, different treatments to include the heavy quark masses into account 
%to calculate the coefficient functions have been used such as the ones used by global fits MSTW08\cite{tr} and CTEQ \cite{acot}, 
% These are based on the general mass variable flavour scheme, however  the fixed flavour number of scheme
%has been tested too \cite{qcdnum}.

%In addition, HERA has addressed the issues related to the PDF uncertainties. So that, 
The uncertainties at HERA are classified in three categories: experimental, model, and parametrisation uncertainties.
The consistency of the input data set 
and its small systematic uncertainties enable us 
to calculate the experimental uncertainties on the PDFs using the 
$\chi^2$ tolerance $\Delta\chi^2=1$. 
%
%The role of correlated   
%systematic uncertainties is no longer crucial since these uncertainties are 
%relatively small. Therefore, the 110(131) systematic uncertainties for HERAPDF1.0 (HERAPDF1.5) 
%are added in quadrature, 
%and the 3 procedural sources of uncertainty are offset, which is the most conservative 
%choice to estimate the experimental uncertainty.
The model uncertainties are evaluated by varying the input assumptions, which are 
the variation of the starting scale and of the $Q^2_{min}$, the variations of the heavy quark masses which are set to the standard values of $m_c=1.4$~GeV and $m_b=4.75$~GeV for the central fit, and the variation of $f_s$.
%The variation of $f_s$ is also considered as a model uncertainty.
The parametrisation uncertainty  
is estimated as an envelope which is formed as a maximal deviation at each $x$ value from the central fit of  10 parameter fits 
with $D$ and $E$ non-zero from Equation~\ref{eqn:pdf}.

%
%The resulting  $\chi^2$ per degree of freedom for the central fit is
%found to be  $574/582$.

%%%%%%%%%%%%%%%%%%%%%%%%%%%%%%%%%%%%%%%%%%%%%%%%%%%%%%%
%%%%%%%%%%%%%%%%%%%%%%%%%%%%%%%%%%%%%%%%%%%%%%%%%%%%%%%
%%%%%%%%%%%%%%%%%%%%%%%%%%%%%%%%%%%%%%%%%%%%%%%%%%%%%%%

\section{Results and Comparisons}\label{sec:results}
\begin{figure}[h!]
\centering
\includegraphics[width=0.495\textwidth]{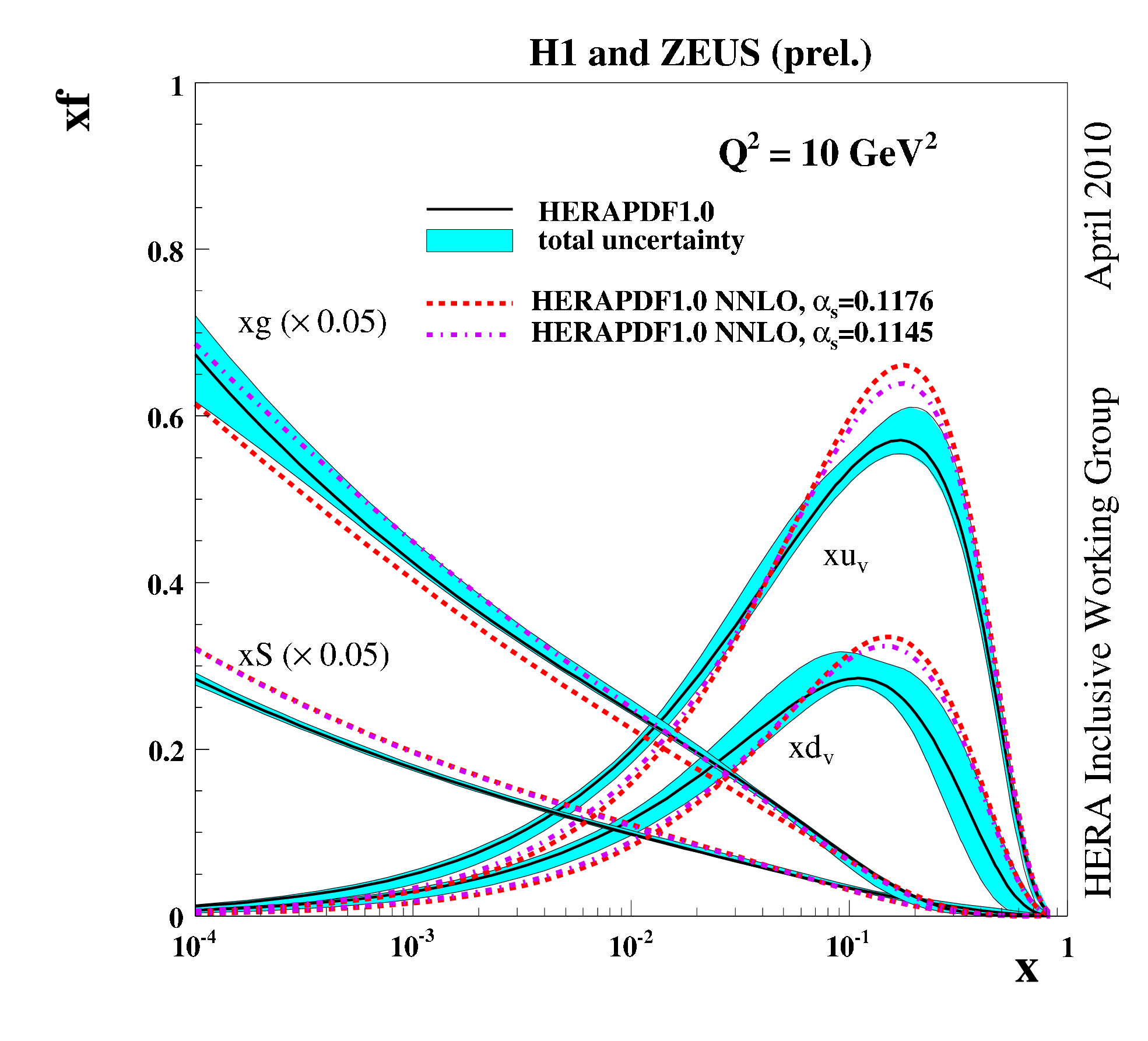}
\includegraphics[width=0.495\textwidth]{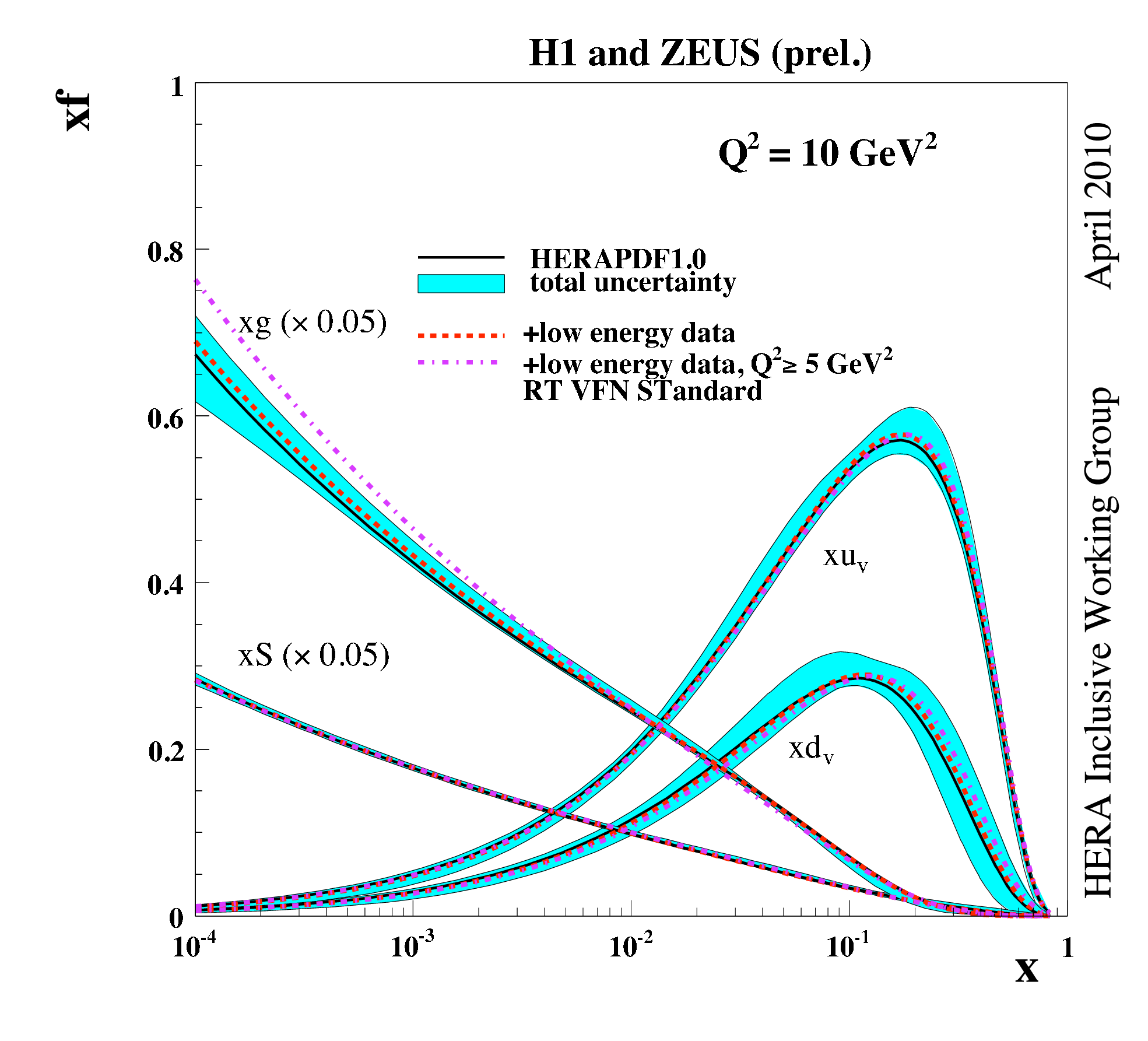}
\caption{Figure shows on the left hand side
the summary plot at the $Q^2=10$ GeV$^2$  with gluon, sea  (which are scaled by a factor of 0.05) and the valence distributions for HERAPDF1.0 at NLO (band) compared illustratively to NNLO fits using $\alpha_S(M_Z)=0.1145$ (dashed) and $\alpha_S(M_Z)=0.1176$ (dotted) lines. 
 On the right hand side, it is shown the comparison between the HERAPDF1.0  (band)  based on HERA I data  and 
 fits including the HERA II of lower energy proton beams with the kinematic cut variation $Q^2\ge 3.5$ (dotted) and  $Q^2\ge 5.0$ (dashed).} 
 \label{Fig:1}
\end{figure}

\vspace*{-0.3cm}
 The NLO QCD analysis has been performed first to the final HERA I data resulting into HERAPDF1.0 which has been published \cite{herapdf} and it will be used as a reference for the new studies. 
The NNLO fit results have been also performed using the same scheme as used for MSTW PDF sets, for different values of the strong coupling,  $\alpha_s(M_Z) =  0.1176$ and $\alpha_s(M_Z) =  0.1145$ \cite{grid}.  Fit results are shown in Figure~\ref{Fig:1}. However, the NNLO fits do not bring improvement in terms of the fit quality with respect to NLO fits
(worse by about 65 and 50 units of $\chi^2$, respectively) for both $\alpha_S$ cases, with a preference for lower value of the strong coupling at NNLO. 

%The kinematic cut variation proved to reveal interesting dependencies for fits including the new preliminary data of lower proton energy data and the results of this study is presented in the next section.

The inclusion of the new preliminary data at low proton beam energy in the HERA QCD fits results in PDF distributions  that agree well with the HERAPDF1.0, as shown in Figure~\ref{Fig:1}. However, a  large sensitivity has been observed when the variation of the kinematic cut has been studied, i.e. $Q^2_{min}>5$~GeV$^2$, which yielded a different PDF shape for the gluon distribution with respect to the the central fit which uses  $Q^2_{min}>3.5$~GeV$^2$.
% This cut variation has been included in the HERAPDF1.0 and is incorporated in the uncertainty band.

\begin{figure}[h!]
\centering
\includegraphics[width=0.495\textwidth]{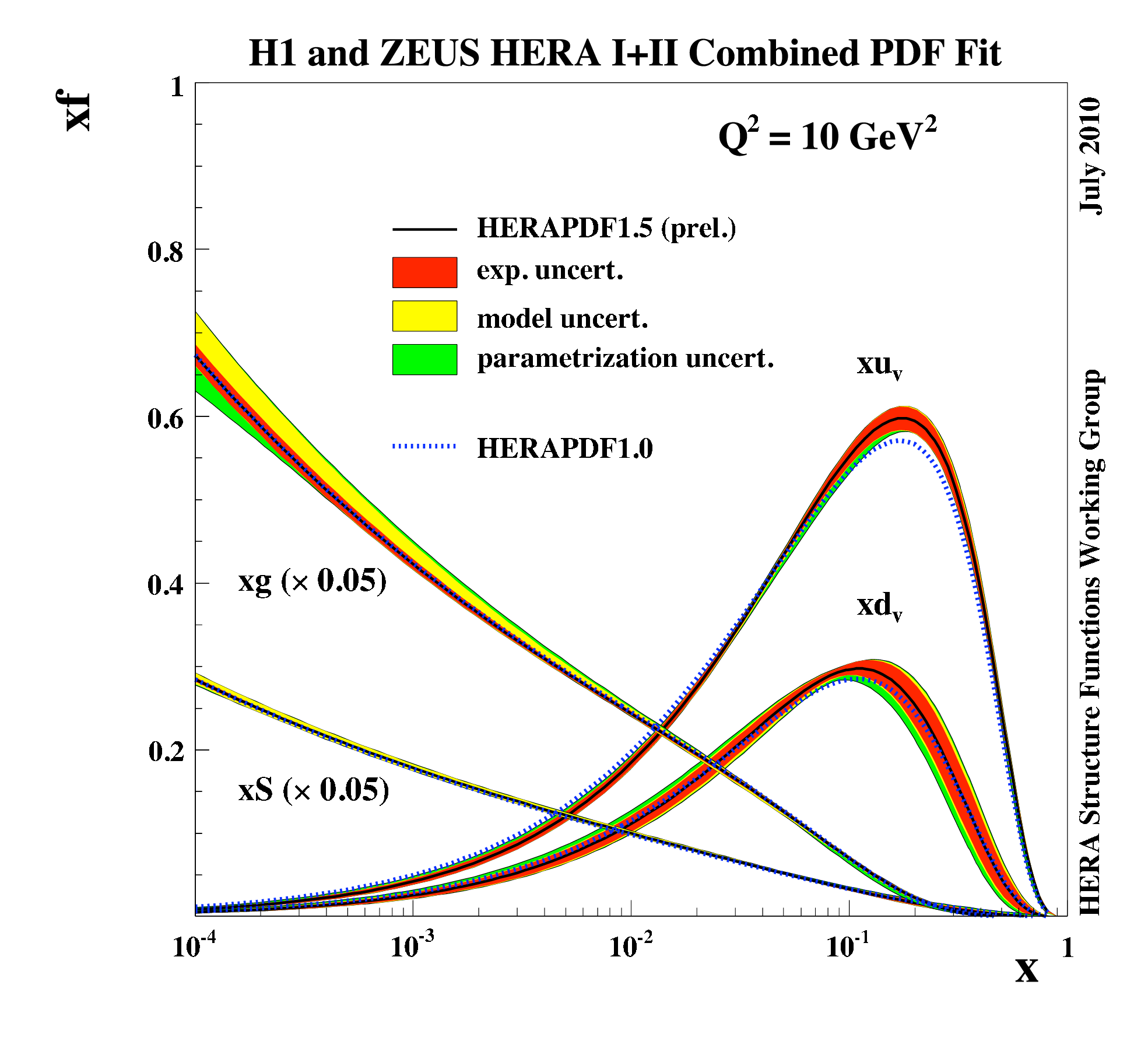}
\includegraphics[width=0.495\textwidth]{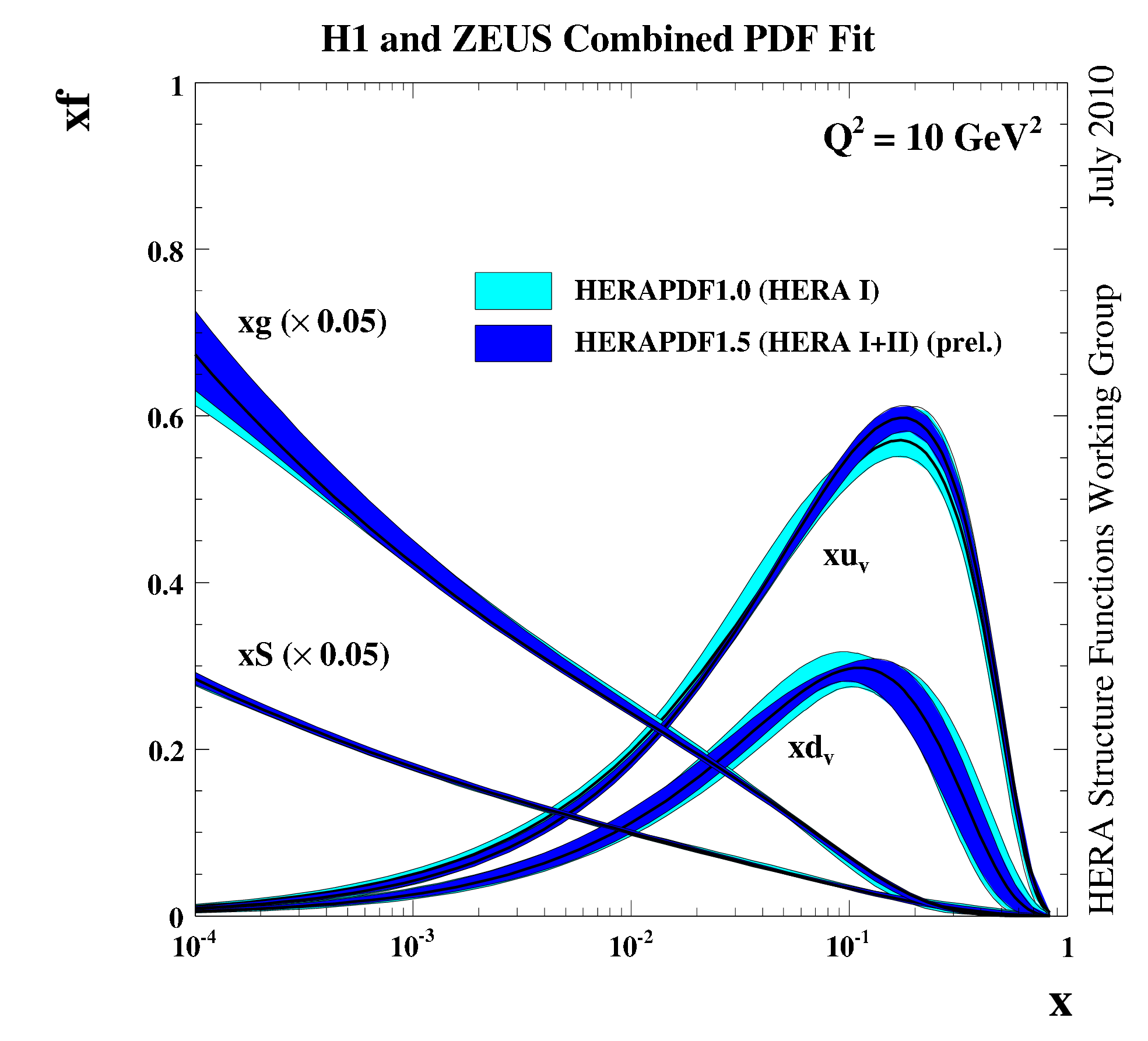}
\caption{Figure shows on the left hand side
the summary plot for the HERAPDF1.5 at the $Q^2=10$ GeV$^2$  with gluon, sea  (which are scaled by a factor of 0.05) and the valence distributions. The errors include the experimental (red), model (yellow) and the PDF parametrisation (green) uncertainties.
 On the right hand side, it is shown the comparison between the HERAPDF1.0  (light color)  based on HERA I data  and HERAPDF1.5 (dark color) based on HERA I and  II data, using total uncertainty band at $Q^2=10$ GeV$^2$.}
 \label{Fig:2}
\end{figure}

The QCD fit analysis of the combined HERA-I inclusive deep inelastic cross sections has been extended to include 
combined HERA II measurements at high $Q^2$ resulting into HERAPDF1.5 \cite{grid}. Figure~\ref{Fig:2} shows that the precision 
of the PDFs at high-$x$ is considerably improved, not only for the experimental uncertainties, but also for the parametrisation uncertainty - particularly in the valence sector, when compared to HERAPDF1.0. This leads to more precise predictions for the LHC process.

The predictions based on HERAPDFs from the DIS process agree well with the  Tevatron jet production, $Z$ and $W$ cross sections  from the $p\bar p$ process and provide a competitive prediction for the LHC $pp$ processes. 
% The effect of including these data on the determination of parton 
%distribution functions is shown in Figure~\ref{Fig:2}, using fits similar to those performed for HERAPDF1.0. The precision of the 
%PDFs at high-$x$ is considerably improved - particularly in the valence sector.

%\vspace*{-3cm}


\begin{thebibliography}{99}
\vspace*{-0.25cm}\bibitem{h1pdf} H1 Collaboration, F.~Aaron \it et al. \em Eur.Phys.J.C64 (2009) 561, arXiv 0904.0929.
\vspace*{-0.25cm}\bibitem{herapdf} H1 Collaboration, ZEUS Collaboration, F. Aaron \it et al. \em, JHEP 1001, 109 (2010), arXiv:0911.0884. 
\vspace*{-0.25cm}\bibitem{pdg} C. Amsler \it et al. \rm (Particle Data Group), Phys. Lett. \bf B667, \rm (2008).
\vspace*{-0.25cm}\bibitem{tr} R.~S.~Thorne code, revised in 2008.
\vspace*{-0.25cm}\bibitem{qcdnum} QCDNUM package, M.~Botje, (2010),  arXiv:1005.1481, \footnotesize \verb$http://www.nikef.nl/h24/qcdnum/index.html$\small
\vspace*{-0.25cm}\bibitem{grid}  The LHAPDF grid files are located at \footnotesize
 \verb$https://www.desy.de/h1zeus/combined$\_\verb$results/index.php?do=proton$\_\verb$structure$ 
 

\end{thebibliography}
\end{document}